\documentclass{emulateapj}

\shorttitle{} \shortauthors{}

\begin{document}

\title{Could FRB 131104 Originate from the Merger of Binary Neutron Stars?}

\author{Z. G. Dai$^{1,2}$, J. S. Wang$^{1,2}$, \& X. F. Wu$^{3,4}$}

\affil{$^1$School of Astronomy and Space Science, Nanjing University, Nanjing 210093, China; dzg@nju.edu.cn\\
$^2$Key Laboratory of Modern Astronomy and Astrophysics (Nanjing University), Ministry of Education, China\\
$^3$Purple Mountain Observatory, Chinese Academy of Sciences, Nanjing 210008, China\\
$^4$Joint Center for Particle, Nuclear Physics and Cosmology, Nanjing
University-Purple Mountain Observatory, Nanjing 210008, China}

\newcommand{\be}{\begin{equation}}
\newcommand{\ee}{\end{equation}}
\newcommand{\g}{\gamma}
\def\ba{\begin{eqnarray}}
\def\ea{\end{eqnarray}}
\def\fv{{F_{\nu}}}
\def\fvm{{F_{\nu,{\rm max}}}}
\newcommand{\e}{\epsilon}

\begin{abstract}
Recently, \cite{DeLaunay2016} discovered a gamma-ray transient, Swift J0644.5$-$5111, associated with the fast radio burst (FRB) 131104. They also reported follow-up broadband observations beginning two days after the FRB and provided upper limits on a putative afterglow of this transient. In this paper, we show that if such a transient drives a relativistic shock as in a cosmological gamma-ray burst (GRB), these upper limits are consistent with an environment of which density is much less than that of an interstellar medium but typical for the outskirts' density of a galaxy when the typical values of three microphysical parameters of the shock are taken. This appears to be inconsistent with the catastrophic event models in which the central engine of Swift J0644.5$-$5111 is surrounded by an interstellar medium, but together with the properties of the gamma-ray transient, favors the binary neutron star merger origin. We further constrain the physical parameters of the postmerger object by assuming that Swift J0644.5$-$5111 results from internal dissipation of a spinning-down pulsar wind, and we find that the postmerger object is an ultra-strongly magnetized, very rapidly rotating pulsar. This merger event should have given birth to a gravitational wave burst, an FRB, and a short GRB or an extended X-ray/$\gamma$-ray emission if a relativistic jet of the GRB is missed. Such ``triplets'' would be testable in the near future with the advanced LIGO and Virgo gravitational-wave observatories.
\end{abstract}


\section{Introduction}

Fast radio bursts (FRBs) are millisecond-duration explosions with emission frequency of Giga-Hertz
\citep{Lorimer2007,Keane2012,Thornton2013,Burke2014,Spitler2014,Spitler2016,Ravi2015,Petroff2015,Champion2015,Masui2015,Keane2016,Ravi2016}.
Most of them have been discovered to arise from high Galactic latitudes and their detected dispersion measures are in the range
of a few hundreds to few thousands parsecs/cm$^3$. These observations strongly suggest that they are
at extragalactic or even cosmological distances. Many origin models have been proposed,
including (1) non-catastrophic, repeatable events, e.g. giant flares from magnetars \citep{Popov2010,Kulkarni2014,Katz2016},
giant pulses from young pulsars \citep{Connor2016,Cordes2016}, eruptions of nearby flaring stars \citep{Loeb2014},
collisions between neutron stars and asteroids/comets \citep{Geng2015,Dai2016a}, and planetary companions around pulsars \citep{Mottez2014},
and (2) catastrophic, non-repeatable events, e.g. mergers of binary compact objects \citep{Totani2013,Kashiyama2013,Mingarelli2015,Wang2016,Zhang2016a,Zhang2016b,Liu2016},
collapses of supra-massive neutron stars to black holes \citep{Falcke2014,Zhang2014}, and giant-glitch-induced events
via the Schwinger mechanism in young magnetars \citep{Lieu2016}.

FRBs seem to fall into two classes: non-repeating and repeating. Only one repeating case discovered so far is FRB 121102,
from direction of which 16 additional bright bursts were detected \citep{Spitler2016,Scholz2016}. This FRB
is inconsistent with all of the catastrophic event models mentioned above. More importantly, it is unlikely to be consistent with
the popular model based on giant pulses from young pulsars. The reasons for this argument are that
giant pulses from the well-known Crab pulsar only have an average temporal width around one microsecond
and that each of such giant pulses has an energy release $\sim 10^{28}\,$erg \citep{Majid2011}.
The former parameter (pulse width) is about three orders of magnitude less than the duration of an FRB while the latter
parameter (pulse energy) is at least ten orders of magnitude smaller than the energy release of a cosmological FRB.
\cite{Dai2016a} proposed an alternative model, in which a repeating FRB may originate from a highly magnetized pulsar
encountering an asteroid belt in another stellar system. They showed that this model can account for all of the properties of an FRB,
including the duration distribution, luminosity, and repeating rate of the 17 bursts of FRB 121102. An advantage of this model is
that repeating bursts from an FRB are expected to provide a unique probe of the physical properties (e.g. size distribution)
of extragalactic asteroids.

Any counterpart to a non-repeating FRB is of great interest and of particular importance for studying the FRB's origin.
\cite{Keane2016} first claimed to detect both a radio afterglow of FRB 140518 and an associated host galaxy.
However, the radio emission was subsequently found to be due to a flaring active galactic nucleus
\citep{Williams2016,Vedantham2016}. Recently, a gamma-ray transient, Swift J0644.5$-$5111, was discovered to be
associated with FRB 131104 at the $3.2\sigma$ confidence level \citep{DeLaunay2016}.
The transient has a duration of $T_{90}\simeq 377\pm 24$\,s ($1\sigma$) and a fluence of
$S_\gamma\simeq (4.0\pm 1.8)\times 10^{-6}\,{\rm erg}\,{\rm cm}^{-2}$.
The detected dispersion measure of this FRB is $779.0\pm0.2$\,pc/cm$^{-3}$ \citep{Ravi2015},
implying that the source is at a cosmological distance $D_L\simeq 3.2$\,Gpc and redshift $z\simeq 0.55$
if the dispersion measure largely arises from an intergalactic medium \citep{Murase2016a}. Given this distance,
the isotropic-equivalent energy output in the gamma-ray band is $E_{\gamma}\simeq 5\times10^{51}$ergs,
which is much larger than the brightest flare of the soft gamma-ray repeater 1806$-$20
\citep{Hurley2005} but typical for the energy release of a cosmological gamma-ray burst (GRB). Thus, this gamma-ray transient
appears to be a long-duration ($T_{90}>100\,$s) GRB or an extended X-ray/$\gamma$-ray emission after a GRB
if a relativistic jet of the GRB does not point to us. The follow-up observations of Swift
and VLT beginning two days after the FRB provide upper limits on a putative broadband afterglow of Swift J0644.5$-$5111,
$4\times 10^{-14}\,{\rm erg}\,{\rm cm}^{-2}\,{\rm s}^{-1}$, $32\,\mu$Jy, and $4.5\,\mu$Jy,
in the X-ray, ultraviolet, and optical bands, respectively \citep{DeLaunay2016}. Very recently,
\cite{Shannon2016} reported their follow-up radio observations beginning three days
after FRB 131104 with the Parkes radio telescope and provided an upper limit ($70\,\mu{\rm Jy}$) on the radio afterglow flux.
Thus, these follow-up observations do not reveal any bright afterglow associated with the gamma-ray transient.

In this paper, we constrain the environmental density of FRB 131104 based on the
follow-up broadband observations of Swift, and show that its environment is typical for the outskirts of a galaxy.
We discuss implications of this constraint as well as the properties of the gamma-ray transient
and suggest that FRB 131104 may originate from the merger of binary neutron stars (BNSs). We also constrain the physical parameters of
the postmerger compact object. Recently, \cite{Murase2016b} estimated an upper limit on the medium density based on
the follow-up radio observations and discussed implications for a relativistic outflow\footnote{While this paper
was close to completion, \cite{Gao2016}, based on the radio data, also gave an upper limit on the medium density of FRB 131104
and discussed the FRB's origin independently. As shown in Section 2, however, the upper limit estimated from the radio data
is much larger than our result.}. We see that our constraint on the medium density is much more stringent than that of \cite{Murase2016b}
if the typical values of three microphysical parameters of a relativistic shock are taken.

This paper is organized as follows. In Section 2, we constrain the environmental density of FRB 131104
by adopting the upper limits on a putative afterglow from Swift. We next discuss the other implications
of FRB131104/Swift J0644.5$-$5111 in Section 3. We find that the BNS merger model
is consistent with both the required very-low-density environment and the physical properties of Swift J0644.5$-$5111.
Finally, we present our conclusions in Section 4.

\section{Upper Limits on the Medium Density}

We assume that a relativistic shock driven by Swift J0644.5$-$5111 expands in a uniform medium with a density of $n$.
In the standard GRB afterglow model \citep{Meszaros1997,Sari1998}, the synchrotron radiation from most of shock-accelerated electrons
with a power-law spectral index of $p$ undergoes two regimes, fast cooling and slow cooling. We further assume
that $E$ is the initial kinetic energy of the shock, and $\e_B$ and $\e_e$ are the fractions of the post-shock energy
density in a magnetic field and electrons, respectively. The transition time $t_0$ of the two regimes depends
on the minimum synchrotron frequency and the cooling frequency \citep{Granot2002},
\begin{eqnarray}
\nu_m & = & 3.7\times 10^{12}(p-0.67)(1+z)^{1/2}\nonumber \\ & & \times
\xi_e^2\epsilon_{e,-1}^2\epsilon_{B,-2}^{1/2}E_{52}^{1/2}t_{\rm day}^{-3/2}\,{\rm Hz}, \label{vm}
\end{eqnarray}
and
\begin{eqnarray}
\nu_c & = & 6.4 \times 10^{18}(p-0.46)e^{-1.16p}\nonumber \\ & & \times
(1+z)^{-1/2}\epsilon_{B,-2}^{-3/2}E_{52}^{-1/2} n_{-2}^{-1}t_{\rm day}^{-1/2}\,{\rm Hz},    \label{vc}
\end{eqnarray}
where $\xi_e\equiv (p-2)/(p-1)$, $t_{\rm day}=t/{\rm day}$, and the other physical quantities $Q_x=Q/10^x$ in cgs units are adopted.
From equations (\ref{vm}) and (\ref{vc}), we have $t_0=1.1\times 10^{-6}\epsilon_{e,-1}^2\epsilon_{B,-2}^2E_{52}
n_{-2}[(1+z)/1.55]\,$days, which implies that the shock will enter the slow cooling regime very rapidly after its formation,
if $p=2.4$ and $n_{-2}<1$. Equation (\ref{vc}) suggests that the cooling frequency will cross the X-ray band in about one day
after the gamma-ray transient. The observed peak flux density of the synchrotron radiation is given by \citep{Granot2002}
\begin{equation}
\fvm = 99(p+0.14)(1+z)\epsilon_{B,-2}^{1/2} E_{52}n_{-2}^{1/2}D_{L,28}^{-2}\,\mu{\rm Jy}\label{fvmax},
\end{equation}
so the flux density at $\nu_m<\nu<\nu_c$ can be written as $F_\nu=\fvm (\nu/\nu_m)^{-(p-1)/2}$, that is,
\begin{eqnarray}
F_\nu & = & 12\nu_{14}^{-(p-1)/2}\epsilon_{e,-1}^{p-1}\epsilon_{B,-2}^{(p+1)/4}E_{52}^{(p+3)/4}
n_{-2}^{1/2}\nonumber \\ & & \times
t_{\rm day}^{-3(p-1)/4}\left(\frac{1+z}{1.55}\right)^{(p+3)/4}\left(\frac{D_L}{3.2{\rm Gpc}}\right)^{-2}\,\mu {\rm Jy}\label{fv},
\end{eqnarray}
where the coefficient is obtained for $p=2.4$. From the upper limit on an X-ray afterglow at $t=2\,$days given by Swift/XRT
\citep{DeLaunay2016}, it is required that
\begin{equation}
\int_{\nu_{X_1}}^{\nu_{X_2}}F_{\nu}d\nu\lesssim 4\times 10^{-14}\,{\rm erg}\,{\rm cm}^{-2}\,{\rm s}^{-1}\label{Xray},
\end{equation}
where $\nu_{X_1}=7.25\times 10^{16}\,$Hz and $\nu_{X_2}=2.41\times 10^{18}\,$Hz.
A combination of equations (\ref{fv}) and (\ref{Xray}) leads to an upper limit on the medium density,
\begin{eqnarray}
n & \lesssim & 2.6\times 10^{-4}\epsilon_{e,-1}^{-2(p-1)}\epsilon_{B,-2}^{-(p+1)/2}E_{52}^{-(p+3)/2}\nonumber \\ & & \times
\left(\frac{1+z}{1.55}\right)^{-(p+3)/2}\left(\frac{D_L}{3.2{\rm Gpc}}\right)^4\,{\rm cm}^{-3}\label{limit1},
\end{eqnarray}
where the coefficient is also given for $p=2.4$. Although Swift/UVOT also provides 6
upper limits on an ultraviolet/optical afterglow of Swift J0644.5$-$5111, we find that the most stringent constraint on the
medium density results from Swift/XRT observations.

We next constrain the medium density from the radio observations. The synchrotron self-absorption frequency is estimated by \citep{Granot2002}
\begin{eqnarray}
\nu_{sa} & = & 3.1\times 10^8\left(\frac{p-1}{3p+2}\right)^{3/5}(1+z)^{-1}\nonumber \\ & & \times
\xi_e^{-1}\epsilon_{e,-1}^{-1}\epsilon_{B,-2}^{1/5}n_{-2}^{3/5}E_{52}^{1/5}\,{\rm Hz}\label{nusa},
\end{eqnarray}
which is typically less than 1\,GHz. Thus, the radio flux density is calculated by $F_\nu=(\nu/\nu_m)^{1/3}\fvm$, that is,
\begin{eqnarray}
F_\nu & = & 46\nu_9^{1/3}\epsilon_{e,-1}^{-2/3}\epsilon_{B,-2}^{1/3}E_{52}^{5/6}n_{-2}^{1/2}t_{\rm day}^{1/2}\nonumber \\ & & \times
\left(\frac{1+z}{1.55}\right)^{5/6}\left(\frac{D_L}{3.2{\rm Gpc}}\right)^{-2}\,\mu {\rm Jy}\label{fvsa}.
\end{eqnarray}
At $\nu=5.5\,$GHz and $t=3\,$days, it is required that $F_\nu\lesssim 70\,\mu$Jy \citep{Shannon2016}, which,
together with equation (\ref{fvsa}), leads to
\begin{eqnarray}
n & \lesssim & 7.7\times 10^{-3}\epsilon_{e,-1}^{4/3}\epsilon_{B,-2}^{-2/3}E_{52}^{-5/3}\nonumber \\ & & \times
\left(\frac{1+z}{1.55}\right)^{-5/3}\left(\frac{D_L}{3.2{\rm Gpc}}\right)^4\,{\rm cm}^{-3}\label{limit2},
\end{eqnarray}
which is much less stringent than the constraint given by equation (\ref{limit1}) for typical values of the model parameters.

We can see from equation (\ref{limit1}) that the upper limit on the medium density is dependent on several physical parameters
of the shock, of which three microphysical parameters ($\epsilon_e$, $\epsilon_B$, and $p$) are very difficult to infer from
the first principle. However, they have been found through
modeling of the broadband emission of a few GRB afterglows \citep{Pan2002}. If $p\simeq 2.4$, $\epsilon_e\sim 0.1$,
$\epsilon_B\sim 0.01$, $E_{52}\sim 1$, $z\simeq 0.55$, and $D_L\simeq 3.2\,$Gpc are typically taken, for example, we have
$n \lesssim 2.6\times 10^{-4}\,{\rm cm}^{-3}$. This upper limit is much smaller than the typical density of
an interstellar medium around long-duration GRBs \citep{Pan2002} but typical for the environmental densities of dark X-ray afterglows
of short-duration GRBs \citep[for reviews see][]{Nakar2007,Berger2014}.

\section{Constraining the Postmerger Object}

In Section 2, we have shown that the environment of FRB 131104 could be similar to those of dark afterglows
of short-duration GRBs. This suggests that FRB 131104 may originate from the BNS merger, because
the BNS system with an average proper velocity of a few hundreds kilometers per second, before its merger,
may spend Giga-years to leave far from their birth site to the outskirts of their host galaxy.
Usually, the typical density of the outskirts is as low as $10^{-4}\,{\rm cm}^{-3}$ \citep{Nakar2007,Berger2014},
in which medium the BNSs will eventually inspiral with each other. The final inspiral before the merger
has been argued to result in an FRB via some physical mechanisms \citep{Totani2013,Wang2016,Zhang2016b}.
In the unipolar inductor model \citep{Wang2016}, for example, if one of the two neutron stars is highly magnetized,
an electric field is induced on another neutron star to accelerate electrons to ultra-relativistic energies instantaneously.
Coherent curvature radiation from these electrons gives rise to an FRB.
If the subsequent merger leads to a black hole surrounded by a neutrino-dominated accretion disk, then this disk will be accreted by
the black hole in a timescale of
\begin{eqnarray}
t_{\rm acc} & \simeq & 0.2\left(\frac{\alpha}{0.1}\right)^{-1}\left(\frac{M_{\rm BH}}{3M_\odot}\right)^{13/7}\nonumber \\ & & \times
\left(\frac{M_{\rm d}}{0.1M_\odot}\right)^{-2/7}\left(\frac{R_{\rm d}}{10R_{\rm S}}\right)^{3/2}\,{\rm s}\label{tacc},
\end{eqnarray}
where $\alpha$ is the viscosity of the disk's gas, $M_{\rm BH}$ is the black hole mass, $M_{\rm d}$ is the disk mass, $R_{\rm d}$
is the outer radius of the disk, and $R_{\rm S}$ is the Schwarzschild radius of the black hole \citep{Narayan2001}.
This timescale can well explain the temporal features of short-duration GRBs but cannot account for the long duration
of Swift J0644.5$-$5111. Therefore, we suggest that the BNS merger as the origin
of FRB 131104/Swift J0644.5$-$5111 may lead to a millisecond magnetar. Such a stable pulsar is possible
if the equation of state for neutron star matter is very stiff \citep[for a discussion see][]{Dai2006}. Its existence
is also supported by general relativistic magnetohydrodynamic simulations of BNS mergers \citep{Gia2013}. Subsequently,
differential rotation of the pulsar's interior will possibly give rise to magnetic reconnection-driven explosive events
\citep[e.g. X-ray flares after short GRBs,][]{Dai2006}, while injection of the stellar rotational energy to a postburst
relativistic shock will perhaps cause plateaus in GRB afterglow light curves, 
as predicted by \cite{Dai1998a,Dai1998b}. In view of this energy injection effect, \cite{Rowlinson2013} collected 
a sample for X-ray afterglow plateaus of short GRBs and provided interesting constraints on the physical parameters of central magnetars.

We next constrain the physical parameters of the postmerger magnetar from the observations on Swift J0644.5$-$5111.
Some short GRBs present an extended emission in the X-ray and gamma-ray range \citep[for a review see][]{Berger2014}. In this paper,
we suggest that Swift J0644.5$-$5111 may be just this emission, which is due to the magnetar spin-down.
Now let's assume that $P$ is the initial period of the magnetar, $B_{\rm s}$ is the stellar surface magnetic field, $I$ is the moment of
inertia, $R_*$ is the magnetar's radius, and $L_{\rm sd}$ is the spin-down luminosity. Owing to internal dissipation mechanisms such as
magnetic reconnection in a pulsar wind \citep[for a discussion see][]{Zhang2013},
the resultant X-ray/$\gamma$-ray transient luminosity is defined to be a fraction $\eta$ of $L_{\rm sd}$, that is,
\begin{eqnarray}
L_\gamma & = & \frac{2\eta}{3c^3}\left(\frac{2\pi}{P}\right)^4B_{\rm s}^2R_*^6\sin^2\theta\nonumber \\
& = & 3.8\times 10^{49}\eta B_{\rm s,15}^2P_{\rm ms}^{-4}R_{*,6}^6\,{\rm erg}\,{\rm s}^{-1}\label{lgamma},
\end{eqnarray}
where $P_{\rm ms}=P/{\rm ms}$ and $\theta=\pi/2$ is the inclination angle between the magnetic dipole moment and the rotation axis.
The initial spin-down timescale is given by
\begin{eqnarray}
T_{\rm sd}\equiv\left(\frac{P}{2\dot{P}}\right)_0=510B_{\rm s,15}^{-2}P_{\rm ms}^2I_{45}R_{*,6}^{-6}\,{\rm s}\label{tsd}.
\end{eqnarray}
Assuming that $L_\gamma=E_\gamma/T_{90}=1.3\times 10^{49}\,{\rm erg}\,{\rm s}^{-1}$ and $T_{\rm sd}=T_{90}=377\,{\rm s}$ for Swift J0644.5$-$5111,
we have
\begin{equation}
P=2.0\eta^{1/2}I_{45}^{1/2}\,{\rm ms}\label{period},
\end{equation}
and
\begin{equation}
B_{\rm s}=2.3\times 10^{15}\eta^{1/2}I_{45}R_{*,6}^{-3}\,{\rm G}\label{bs}.
\end{equation}

We further calculate $\eta$. For non-accreting, rotation-powered pulsars, the X-ray/$\gamma$-ray luminosity is found
to be correlated with the spin-down luminosity \citep{Seward1988,Becker1997,Possenti2002,Cheng2004,Li2008}. This correlation
was recently refitted by \cite{Yi2014}, who gave a power-law expression of
\begin{eqnarray}
L_\gamma= 10^{-13.56\pm 1.90}\left(\frac{L_{\rm sd}}{{\rm erg}\,{\rm s}^{-1}}\right)^{1.28\pm0.05}\,{\rm erg}\,{\rm s}^{-1}\label{lg}.
\end{eqnarray}
We apply this correlation to millisecond magnetars and obtain the X-ray/$\gamma$-ray
emission efficiency,
\begin{eqnarray}
\eta\equiv\frac{L_\gamma}{L_{\rm sd}}= 10^{-10.59\pm 1.54}\left(\frac{L_\gamma}{{\rm erg}\,{\rm s}^{-1}}\right)^{0.22\pm0.03}\label{eta}.
\end{eqnarray}
For $L_\gamma=1.3\times 10^{49}\,{\rm erg}\,{\rm s}^{-1}$, we find that $\eta$ is in the range of 0.04 to 1.0. Please note that
all the values of $\eta>1$ are unphysical and thus have been neglected. From equations (\ref{period}) and (\ref{bs}),
therefore, we rewrite the physical parameters of the postmerger magnetar as
\begin{equation}
P= (0.4\,\,{\rm to}\,\,2.0)\times I_{45}^{1/2}\,{\rm ms}\label{period1},
\end{equation}
and
\begin{equation}
B_{\rm s}= (0.46\,\,{\rm to}\,\,2.3)\times 10^{15}I_{45}R_{*,6}^{-3}\,{\rm G}\label{bs1}.
\end{equation}
We can see that the postmerger object is indeed an ultra-strongly magnetized, very rapidly rotating pulsar.

As shown by \cite{Dai2016b}, if this pulsar is a neutron star and has a sub-millisecond rotation period, its rotational energy
may be quickly lost as a result of some gravitational-radiation-driven instabilities (e.g. the r-mode instability). If such a pulsar
is a strange quark star, however, these instabilities may be highly suppressed, owing to a large bulk viscosity associated with
the nonleptonic weak interaction among quarks, and thus most of the rotational energy could be extracted via magnetic dipole
radiation \citep{Dai2016b}. The long duration and high luminosity of Swift J0644.5$-$5111 suggest that this rotational energy should
have been mainly lost via magnetic dipole radiation, favoring a strange quark star candidate.
This conclusion is not only consistent with the early proposal of \cite{Dai1998b} but also supported by a statistical analysis
of X-ray afterglow plateaus of short GRBs by \cite{Li2016}.

By the way, we discuss why internal dissipation of a free millisecond-magnetar wind can produce an X-ray/$\gamma$-ray emission.
As analyzed by \cite{Zhang2013}, owing to a magnetic reconnection process in the wind, the typical synchrotron photon energy
of accelerated electrons with a power-law distribution is estimated by $\varepsilon_\gamma\simeq 80L_{\gamma,48}^{1/2}
(R_{\rm m}/10^{15}{\rm cm})^{-1}(\eta/0.01)^{3/2}(\sigma/10^4)^2\,$keV, where $R_{\rm m}$ is the radius at which magnetic reconnection
take places, and $\sigma$ is the magnetization parameter at $R_{\rm m}$. In fact, the magnetization parameter of the wind evolves with
radius as $\sigma\propto R^{-1/3}$, which causes a decrease of $\sigma$ from its initial value $\sigma\sim 10^6$ at $R_0\sim 10^7\,$cm
to $\sigma\sim 2\times 10^3$ at $R_{\rm m}\sim 10^{15}\,$cm \citep{Zhang2013}. Combining these typical parameters with
$L_\gamma=1.3\times 10^{49}\,{\rm erg}\,{\rm s}^{-1}$, we find $\varepsilon_\gamma\sim 92(\eta/0.04)^{3/2}\,$keV, which is
in the hard X-ray to soft gamma-ray band. Therefore, we conclude that such an emission leads to Swift J0644.5$-$5111.

\section{Conclusions}
Based on the recent discovery of FRB 131104 associated with the gamma-ray transient Swift J0644.5$-$5111 and the upper limits on a putative afterglow from follow-up broadband observations, in this paper, we have first shown that if this gamma-ray transient drives a relativistic shock as in a cosmological GRB, these upper limits require that the environmental density is much less than that of an interstellar medium but typical for the outskirts' density of a galaxy. This is likely to rule out the catastrophic event models in which the central engine of Swift J0644.5$-$5111 is surrounded by an interstellar medium, but supports the BNS merger origin.

Furthermore, we have constrained the physical parameters of the postmerger compact object. By assuming that Swift J0644.5$-$5111 results from internal dissipation of a spinning-down pulsar wind, together with the long duration, high luminosity, and typical photon energy of the gamma-ray transient, we have found that the postmerger object is an ultra-strongly magnetized, very rapidly rotating pulsar. We have also suggested that this pulsar could be a strange quark star candidate, due to the fact that a newborn rapidly rotating strange quark star does not suffer from some gravitational-radiation-driven instabilities including the r-mode instability.

Finally, what we would point out is the most important implication of the BNS merger origin of FRB 131104, that is, this merger event should have given birth to a gravitational wave burst, an FRB, and a short GRB or an extended X-ray/$\gamma$-ray emission if a relativistic jet of the GRB is missed. Such ``triplets'' would be testable with the advanced LIGO and Virgo gravitational-wave observatories. The other implications include: (1) future observations would be able to test the validity of the BNS merger model and severely constrain the ratio of the BNS merger to FRB rates \citep{Callister2016}, and (2) joint detections of gravitational waves and electromagnetic signals would provide an independent probe of Einstein's weak equivalence principle \citep{Wu2016}.

\acknowledgements
We thank He Gao, Bruno Giacomazzo,  and Bing Zhang for helpful comments. This work was supported by the National Basic Research Program
(``973" Program) of China (grant No. 2014CB845800) and the National Natural Science Foundation of China (grant Nos. 11573014,
11322328 and 11473012). X.F.W. was also partially supported by the Youth Innovation Promotion Association (No. 2011231) and
the Strategic Priority Research Program ``The Emergence of Cosmological Structure'' (grant No. XDB09000000) of the Chinese
Academy of Sciences.

\end{document}